\documentclass[preprint,aps]{revtex4}

\usepackage{graphicx}
\usepackage{dcolumn}
\usepackage{bm}
\usepackage{multirow}
\usepackage{color}
\usepackage{ulem}
\usepackage{amsmath}
\usepackage{gensymb}

\begin{document}
\title{Ultra-low power threshold for laser induced changes in optical properties of 2D Molybdenum dichalcogenides}

\author{Fabian Cadiz$^{1}$}
\author{Cedric Robert$^{1}$}
\author{Gang Wang$^{1}$}
\author{Wilson Kong$^{2}$}
\author{Xi Fan$^{2}$}
\author{Mark Blei$^{2}$}
\author{Delphine Lagarde$^{1}$}
\author{Maxime Gay$^{1}$}
\author{Marco Manca$^{1}$}
\author{Takashi Taniguchi$^{3}$} 
\author{Kenji Watanabe$^{3}$} 
\author{Thierry Amand$^{1}$}
\author{Xavier Marie$^{1}$}
\author{Pierre Renucci$^{1}$}
\author{Sefaattin Tongay$^{2}$}
\author{Bernhard Urbaszek$^{1}$}

\affiliation{%
$^1$ Universit\'e de Toulouse, INSA-CNRS-UPS, LPCNO, 135 Av. Rangueil, 31077 Toulouse, France\\
$^2$ School for Engineering of Matter, Transport and Energy, Arizona State University, Tempe, Arizona 85287, USA\\
$^3$ Advanced Materials Laboratory, National Institute for Materials Science, Tsukuba, Ibaraki 305-0044, Japan}


\begin{abstract}
The optical response of traditional semiconductors depends on the laser excitation power used in experiments. For two-dimensional (2D) semiconductors, laser excitation effects are anticipated to be vastly different due to complexity added by their ultimate thinness, high surface to volume ratio, and laser-membrane interaction effects. We show in this article that under laser excitation the optical properties of 2D materials undergo irreversible changes. Most surprisingly these effects take place even at low steady state excitation, which is commonly thought to be non-intrusive. In low temperature photoluminescence (PL) we show for monolayer (ML) MoSe$_2$ samples grown by different techniques that laser treatment increases significantly the trion (i.e. charged exciton) contribution to the emission compared to the neutral exciton emission. Comparison between samples exfoliated onto different substrates shows that laser induced doping is more efficient for ML MoSe$_2$ on SiO$_2$/Si compared to h-BN and gold. For ML MoS$_2$ we show that exposure to laser radiation with an average power in the $\mu$W/$\mu$m$^2$ range does not just increase the trion-to-exciton PL emission ratio, but may result in the irreversible disappearance of the neutral exciton PL emission and a shift of the main PL peak to lower energy.
\end{abstract}


\maketitle
\textit{\textbf{Introduction.---}}
Atomically thin layers of Van der Waals bonded materials open up new possibilities for physics and chemistry on the nanoscale and for new applications in electronics and photonics \cite{Novoselov:2005a,Geim:2013a,Lopez:2013a,Chhowalla:2013a,Li:2014b,Yu:2015a,Castellanos:2016a}. Here the group-VI Transition metal dichalcogenides (TMDs) of the form MX$_2$, where M=$Mo,W$ and X=$S,Se$, are of particular interest: These indirect semiconductors in bulk form become direct semiconductors when thinned down to one monolayer (ML), which makes them especially attractive for optoelectronics \cite{Pospischil:2014a,Mak:2016a,Sundaram:2013a,Withers:2015a,Liu:2015a}.\\
\indent The optical properties of TMD MLs are most commonly probed in optical spectroscopy: A laser of suitable energy will create an electron-hole pair (exciton) and subsequently the photoluminescence (PL) emission is monitored. This simple technique led to the discovery of ML MoS$_2$ having a direct bandgap \cite{Mak:2010a,Splendiani:2010a,Eda:2011a}. The optical properties are governed by neutral and charged excitons (trions). There are different physical origins of the resident carriers probed in optical spectroscopy: intrinsic dopants, molecules on the ML surface, carriers trapped at the ML-substrate interface \cite{Currie:2015a,Lu:2014a,Mouri:2013a,McDonnell:2014a,Najmaei:2014a}. Here we show that the excitation laser itself can have profound impact on the optical properties, in particular on the doping and emission from localized states.\\
\begin{figure*}
\includegraphics[width=0.7\textwidth]{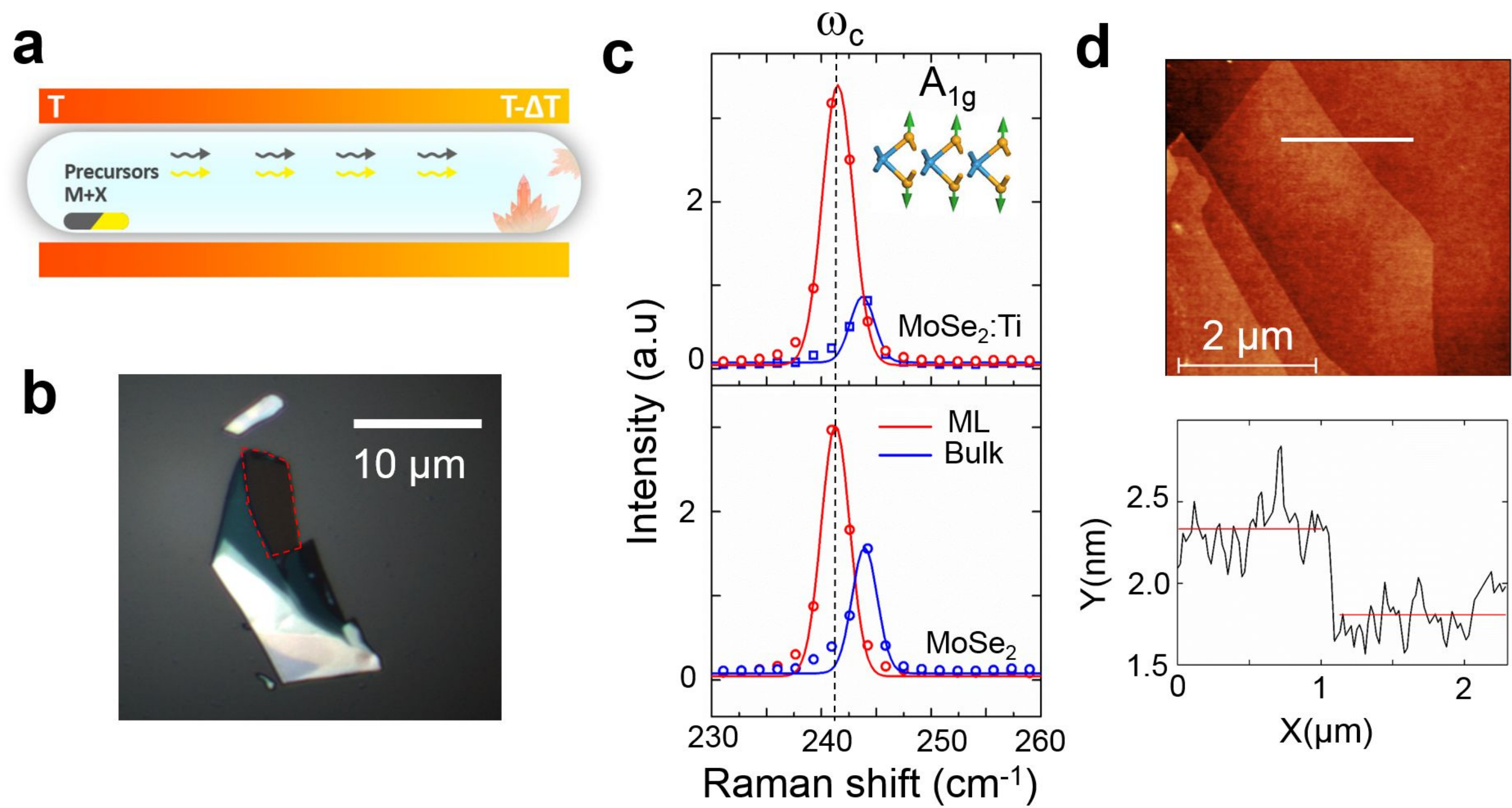}
\caption{\label{fig:fig0} \textbf{MoSe$_2$ monolayer samples} (a) Schematic of chemical vapor phase transport (VPT) set-up used for MoSe$_2$ bulk sample growth (b) Optical microscope image of monolayer MoSe$_2$ (red dotted outline) exfoliated from VPT grown bulk onto SiO$_2$/Si. (c) Raman spectra for MoSe$_2$ Ti doped (top panel) and not intentionally doped (bottom panel) confirm monolayer thickness by comparison with \cite{Tonndorf:2013a} (d) Atomic force microscopy show monolayer step-height.
}
\end{figure*}
\indent We show that this a general phenomenon for MoX$_2$ MLs by systematically comparing MoSe$_2$ ML exfoliated from ultra-pure vapor phase transport (VPT) grown samples with commercial samples. In ML MoSe$_2$ the sharp exciton emission lines at cryogenic temperatures serve as a sensitive probe of the trion-to-neutral exciton PL emission ratio, which indicates presence/absence of additional carriers. As the Ti-doped sample exfoliated from VPT grown bulk is presumably p-type (confirmed in Hall measurements on bulk) an increase of the trion contribution is consistent with adding holes. We then show that in theses MoSe$_2$ sample systems the effect of laser doping is clearest on SiO$_2$ substrates and much weaker in samples exfoliated onto h-BN and Au. Laser induced doping is also studied in ML MoS$_2$. Here exposure to pulsed lasers leads to irreversible changes in the emission spectrum: the neutral exciton signature is lost in PL and the main PL emission is shifted in energy below the initial trion emission. This laser engineering of optical properties for ML MoS$_2$ is demonstrated at T=4~K and at room temperature.\\
\indent \textit{\textbf{MoSe$_2$ and MoS$_2$ monolayers exfoliated from different bulk samples.---}}
Chemical VPT was used to grow MoSe$_2$ and Ti-doped MoSe$_2$ crystals. Molybdenum (99.9999\%), Selenium shots (99.9999\%) and I$_2$, which acted as a transport agent, were sealed in a quartz tube at 5$\times$10$^{-5}$ Torr vacuum. The precursors (hot end) were kept at 1085$^{\circ}$C in a 3-zone horizontal furnace while maintaining a 55$^{\circ}$C temperature difference on the cold end ($\approx$1030$^{\circ}$C) to initiate nucleation and growth \cite{Wildervanck:1970a}, see schematic in Fig.~\ref{fig:fig0}a.  To incorporate Ti into MoSe$_2$, the temperature difference between the two ends of the tube varied from 55 to 65$^{\circ}$C. For comparison bulk MoSe$_2$ crystals were purchased from 2D semiconductors. The MoS$_2$ bulk was also supplied by 2D semiconductors. Using a dry-stamping technique \cite{Gomez:2014a} MLs from different bulk sources were deposited on either SiO$_2$/Si, h-BN \cite{Taniguchi:2007a} or gold substrates. Monolayer thickness was confirmed by several techniques: in optical contrast measurements (Fig.~\ref{fig:fig0}b), in Raman spectroscopy \cite{Tonndorf:2013a} (Fig.~\ref{fig:fig0}c) and atomic force microscopy (Fig.~\ref{fig:fig0}d). The thickness of MoSe$_2$ MLs measures $\sim$0.7~nm in height, and the out-of-plane (A$_{1g}$) peak of MoSe$_2$ softens from bulk (blue dashed line) to ML (red dashed line) as shown in Fig.~\ref{fig:fig0}c due to much reduced restoring forces acting on individual ML sheets.\\
\indent \textit{\textbf{Optical spectroscopy techniques.---}} A purpose-build micro-PL set-up is used to record the PL spectra in the temperature range $T=4-300$~K \cite{Wang:2014b}. The sample is placed on 3-axis stepper motors to control the sample position with nm precision inside the low-vibration closed cycle He cryostat. MLs were excited with a linearly polarized cw laser ($532$ nm or $633$ nm wavelength) or with $1.5$ ps pulses at  $400$ nm generated by a tunable mode-locked  frequency-doubled Ti:Sa laser with a repetition rate of $80$ MHz \cite{Lagarde:2014a}. In all cases, the excitation spot diameter is diffraction limited $\leq1\mu$m, i.e considerably smaller than the ML size of typically $\sim 10~\mu$m$\times10~\mu$m. The PL emission is dispersed in a spectrometer (f=50~cm) and detected with a liquid nitrogen cooled Si-CCD back-illuminated deep-depletion camera. \\
\begin{figure*}
\includegraphics[width=0.8\textwidth]{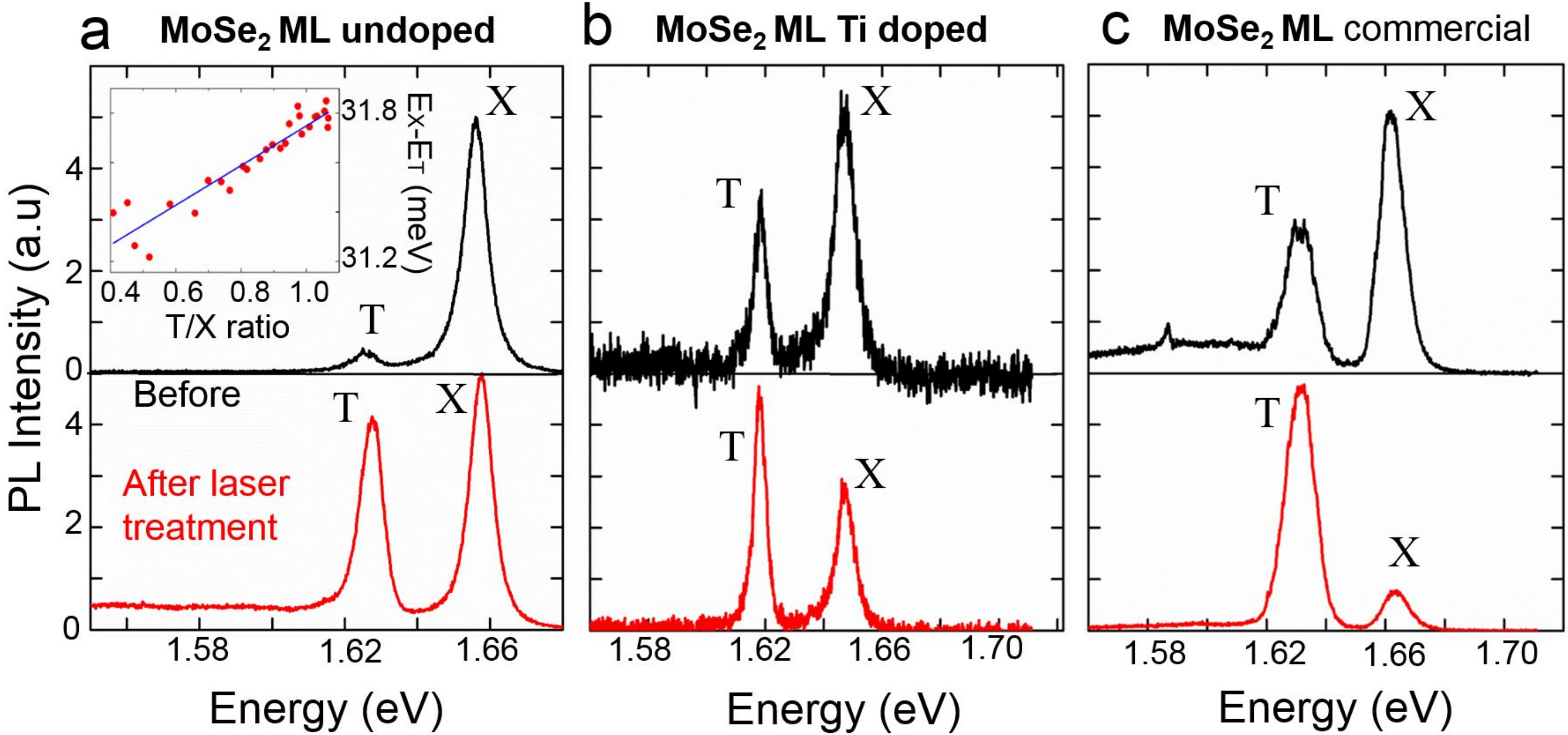}
\caption{\label{fig:fig1} \textbf{Low temperature PL of monolayer MoSe$_2$} (a) PL spectrum of MoSe$_2$ ML (exfoliated from a VPT-grown bulk) at T=8~K for a $40$~nW cw excitation ($633$ nm) before and after being exposed to a power of $40$~$\mu$W at 633~nm during 4 minutes. After exposure, the trion intensity increases. The inset shows that the trion's dissociation energy increases as a function of the T/X ratio, which is a signature of laser-induced doping of the ML. (b) Same as (a) for a Ti-doped MoSe$_2$ ML (exfoliated from a VPT-grown bulk). (c) Same as (a) for MoSe$_2$ ML exfoliated from a commercial (2D Semiconductors) bulk crystal. All samples were mechanically exfoliated on $90$ nm-thick SiO$_2$ layer on top of a Si substrate.
}
\end{figure*}
\indent \textit{\textbf{Effect of Laser radiation on optical properties of MoSe$_2$ MLs}.---}
Changes of the optical emission of ML TMDs as a function of laser excitation power have been reported many times in the literature. Here two different scenarios have to be distinguished. On the one hand, creating more carriers (excitons) will induce eventually interactions between free and localized excitons, trions and resident carriers and possibly result in biexciton formation \cite{Plechinger:2015a,You:2015a,Shang:2015a,Kim:2016a,Kumar:2014b,Singh:2014a,Sun:2014a,Yu:2016a,Sie:2015a}. On the other hand, laser excitation can physically induce non-reversible changes of the ML sample and therefore its optical emission \cite{Currie:2015a,Mitioglu:2013a,He:2016a,Kaplan:2016a,Wang:2016c,Li:2013a,Castellanos-Gomez:2012a}. For intermediate laser power ranges all these physical processes might occur simultaneously, which might explain the wide range of values reported for the neutral and charged exciton PL emission energy in the literature especially for MoS$_2$ and WS$_2$. \\
\begin{figure*}
\includegraphics[width=0.75\textwidth]{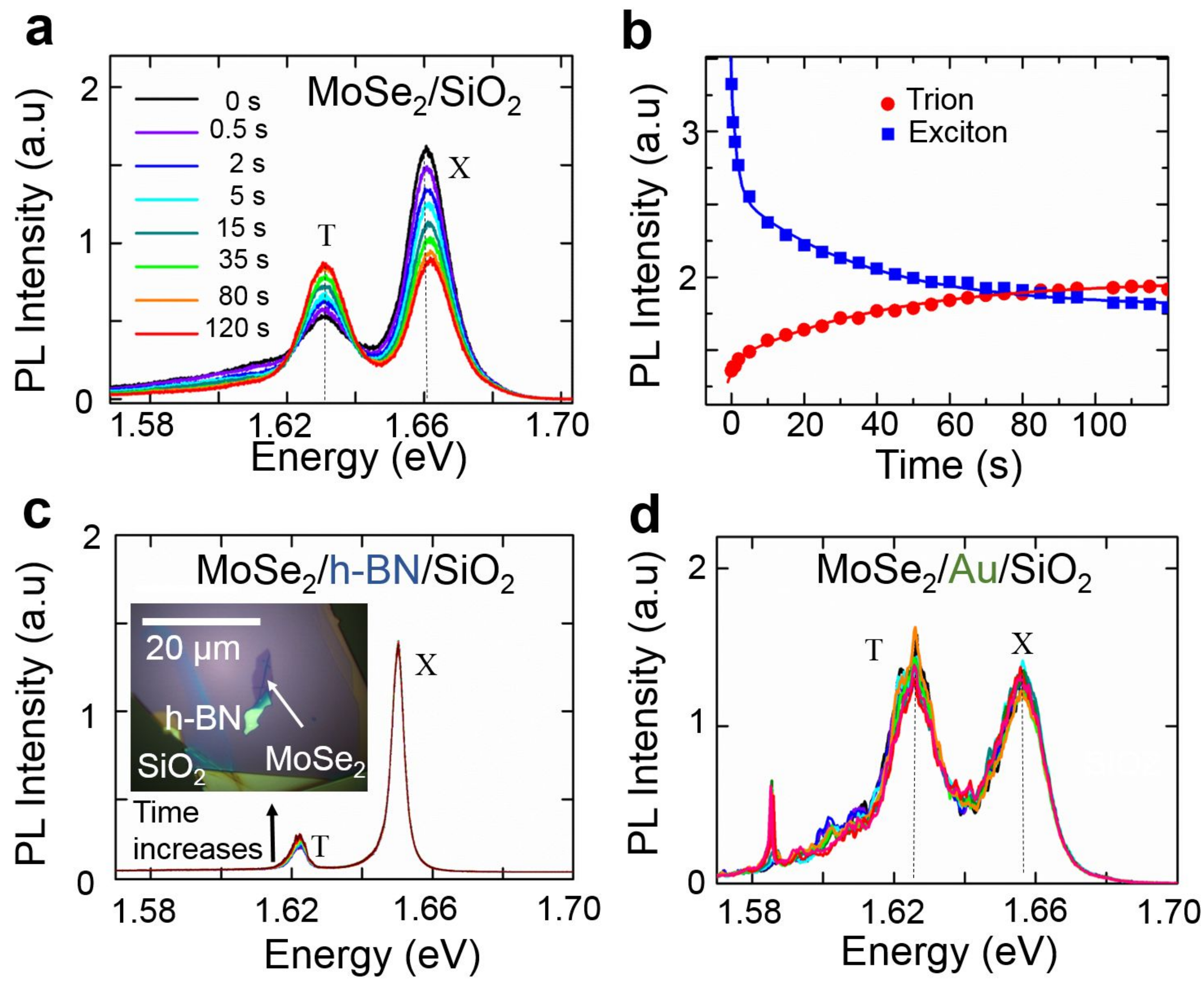}
\caption{\label{fig:fig2} \textbf{Time evolution and substrate effects for laser treatment of ML MoSe$_2$} (a) Time-evolution of the PL spectrum at $T=8$ K of a VPT-grown MoSe$_2$ ML exfoliated onto a SiO$_2$/Si substrate while being exposed to a cw excitation ($532$ nm) at $230$ $\mu$W.  (b) T and X integrated intensity as a function of time for the conditions of panel (a), revealing doping of the ML on a timescale of several minutes. (c)Ti-doped VPT-grown MoSe$_2$ ML exfoliated onto few-layer h-BN. Time-evolution of the PL spectrum at $T=4$~K  while being exposed to a cw excitation ($633$ nm) at $40\; \mu$W during 4 minutes. Only a very slight increase of the T emission is observed. (d) Time-evolution of the PL spectrum at $T=4$~K of a commercial MoSe$_2$ ML exfoliated onto a gold layer (different colours for different times as in panel (a)). The ML is exposed to a cw excitation ($633$ nm) at $50\;\mu$W during 10 minutes. No significant change of the spectrum is observed.  
}
\end{figure*}
\indent Our target is to distinguish between these two scenarios. A simple approach is to perform power dependent measurements several times to verify if the laser radiation induced irreversible changes to the spectra. For this we start each experiment with pristine, as-exfoliated flakes that we probe at very low laser power (nW/$\mu$m$^2$). We start with low temperature measurements at T=10~K in vacuum (10$^{-6}$~mbar) on MoSe$_2$ samples grown under controlled conditions and exfoliated onto different substrates for comparison.\\
\indent Panel (a) of Fig.\ref{fig:fig1} shows the PL spectrum using an extremely low excitation power of $40$~nW (black curve) of ML MoSe$_2$ exfoliated onto SiO$_2$/Si from a VPT grown bulk sample. The PL spectrum of the \textit{pristine} sample is dominated by strong and spectrally narrow (FWHM$\approx8$~meV) neutral exciton emission (X) at 1.66~eV. A much smaller peak attributed to the trion (T) is detected at 1.63~eV. The X and T PL energies are in agreement with standard MoSe$_2$ ML samples \cite{Ross:2013a,Wang:2015a,Wang:2015c}. The PL FWHM for both transitions of just a few meV are among the best reported in the literature and confirm the excellent sample quality of the pristine flakes. After this initial low power measurement, the laser excitation power at this sample position is increased to $40$~$\mu$W and kept constant during 4 minutes (no measurable evolution is detected for longer times). Directly afterwards, the laser power is lowered to $40$~nW, to compare with the black curve before laser treatment, and the PL response is measured (red curve). Remarkably, the trion-to-neutral exciton PL emission intensity ratio $T/X$ has significantly changed, indicating strong doping as a result of the laser treatment. Shown in the inset is the trion's dissociation energy, defined as the difference between the emission energy of the neutral exciton ($E_X$) and the trion ($E_T$), as a function of the $T/X$ PL intensity ratio. In a simple picture, the trion's dissociation energy can be written as \cite{Mak:2013a} :
$$E_X-E_T= E_F+ E_T^B  $$

\noindent
where $E^T_B$ is the trion binding energy (typically $\sim 25$~meV) and $E_F$ is the Fermi level with respect to the bottom of the conduction band for electrons, with respect to the top of the valence band for holes, respectively.  The observed linear increase of $E_X-E_T$ when $T/X$ increases is a signature of an increase of the Fermi level, i.e., of the doping of the ML. Panels (b) and (c) of Fig.\ref{fig:fig1} show the laser-induced doping of Ti-doped VPT grown MoSe$_2$ MLs and commercial MoSe$_2$ MLs, respectively. For the different sample sets compared in Fig.~1a,b and c, exposure to a $40\;\mu$W excitation during 4 minutes produces an irreversible change in the $T/X$ spectral weight. 
We have detected irreversible changes of T/X in vacuum even for powers as low as 1$\mu$W/$\mu$m$^2$. Please note that in many optical spectroscopy measurements of TMDs MLs the excitation densities are much larger than the ones used in this study.\\
\begin{figure*}
\includegraphics[width=0.8\textwidth]{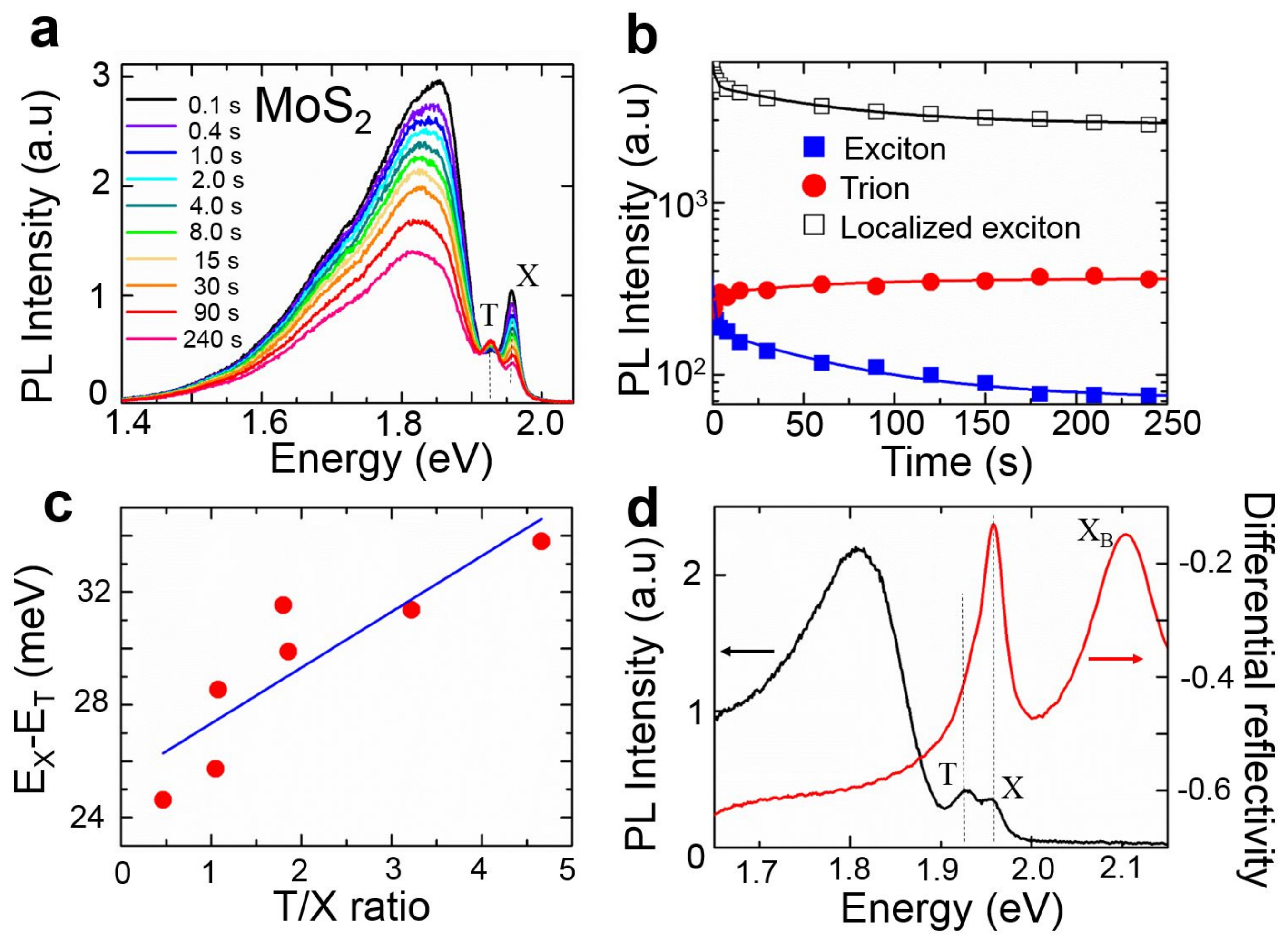}
\caption{\label{fig:fig3} \textbf{cw Laser treatment of ML MoS$_2$} (a) commercial MoS$_2$ ML exfoliated onto a SiO$_2$/Si substrate. Time evolution of the PL spectrum while being exposed to a cw excitation $532$ nm at $60\;\mu$W, revealing X, T and localized exciton emission features which evolve over several minutes. (b) T, X and localized emission integrated intensity as a function of time for the conditions of panel (a). Both the X and the localized exciton emission decrease as a function of time, while the T gains in intensity. (c) Trion's dissociation energy as a function of the T/X ratio for a MoS$_2$ ML. The different values of $T/X$ have been obtained by exposing the ML to different laser powers at different positions on the same ML flake. (d) PL spectrum  at $0.4\mu$~W~cw excitation ($532$~nm) measured after laser treatment of a MoS$_2$ ML, revealing high T emission (black curve). Also shown is the differential reflectivity spectrum (red curve) in which the main absorption is dominated by the X resonance at $1.96$~eV, the additional transition at $\approx$2.1~eV is the B-exciton (X$_B$) \cite{Mak:2010a}.
}
\end{figure*}
\indent Although the doping due to the laser treatment is clearly visible in all panels in Fig.~\ref{fig:fig1}, the nature of the doping (n- or p-type) still needs to be determined. Here the results on the Ti-doped VPT sample give helpful indications in Fig.~\ref{fig:fig1}b: Hall conductivity measurements on the bulk sample before exfoliation indicate p-doping - we therefore assume that the trion in the pristine sample when deposited on SiO$_2$ is positively charged. The T/X ratio increases gradually as the laser power is increased, which would be consistent with extra holes being created by the laser treatment.\\
\indent \textit{\textbf{Dynamics of Laser treatment in ML MoSe$_2$}.---} The dynamics of this laser doping is shown in panel (a) of Fig.~\ref{fig:fig2} in real-time for a commercial MoSe$_2$ ML exposed to a cw excitation at $230\;\mu$W. The evolution of the X and T integrated PL intensity as a function of time is shown in panel (b), where it can be seen that the X intensity presents a first rapid decrease (within seconds) followed by a second slower decrease in a timescale of several minutes. The trion evolution, in contrast, is marked by an increase on similar timescales. Laser treatment for longer than 4 minutes did not result in any further, measurable evolution of the PL spectrum. The total PL intensity (Trion + exciton) is decreasing in Fig.~\ref{fig:fig2}b as more carriers are added, which might be due to the complex interplay between optically bright and dark state of the trion and exciton \cite{Arora:2015b,Echeverry:2016a,Kormanyos:2015a}. \\
\indent \textit{\textbf{Results for ML MoSe$_2$ on different substrates}.---} In order to check the applicability of this laser-induced doping of MoSe$_2$ MLs for different device geometries, we also exfoliated MoSe$_2$ MLs onto a few-layer h-BN film \cite{Taniguchi:2007a} and also a 50 nm- thick gold film. When deposited onto h-BN, only a very small increase of the T emission was observed after 4 minutes exposure to a cw laser at $40\;\mu$W, as shown in panel (c) of Fig.~\ref{fig:fig2}. When deposited on top of a gold layer, no effect of the laser treatment was observed even after $10$ minutes of exposure to a cw laser at $40\;\mu$W. In our experiments the MLs exfoliated onto SiO$_2$/Si showed by far the strongest impact of laser radiation on the optical emission properties. \\
\indent \textit{\textbf{Treatment of ML MoS$_2$ with cw Lasers}.---} The drastic changes of the emission properties of MoSe$_2$ MLs as a function of laser power raise the question if similar effects can be observed in other materials. Below we show our systematic study on ML MoS$_2$ which confirms the strong impact of the excitation laser on the local doping in the layer. Panel (a) of Fig. \ref{fig:fig3} shows the time-evolution of the PL signal from MoS$_2$ MLs at T=8~K when exposed to a cw excitation at $60$~$\mu$W at 532~nm. Three distinct PL emission peaks are observed, associated to the X ($1.96$ eV),T ($1.93$ eV), and spectrally broad localized exciton emission \cite{Cadiz:2016a}, which peaks approximately at $1.85$ eV. Please note that under laser illumination, both the X and the localized emission decrease, whereas the trion increases (panel (b) of Fig.\ref{fig:fig3}). Again, the dynamics is characterized by a fast and a slow component, of several seconds and several minutes, respectively. Also in this case the total PL intensity decrease as for ML MoSe$_2$ in Fig.~\ref{fig:fig2}b. \\
\begin{figure*}
\includegraphics[width=0.8\textwidth]{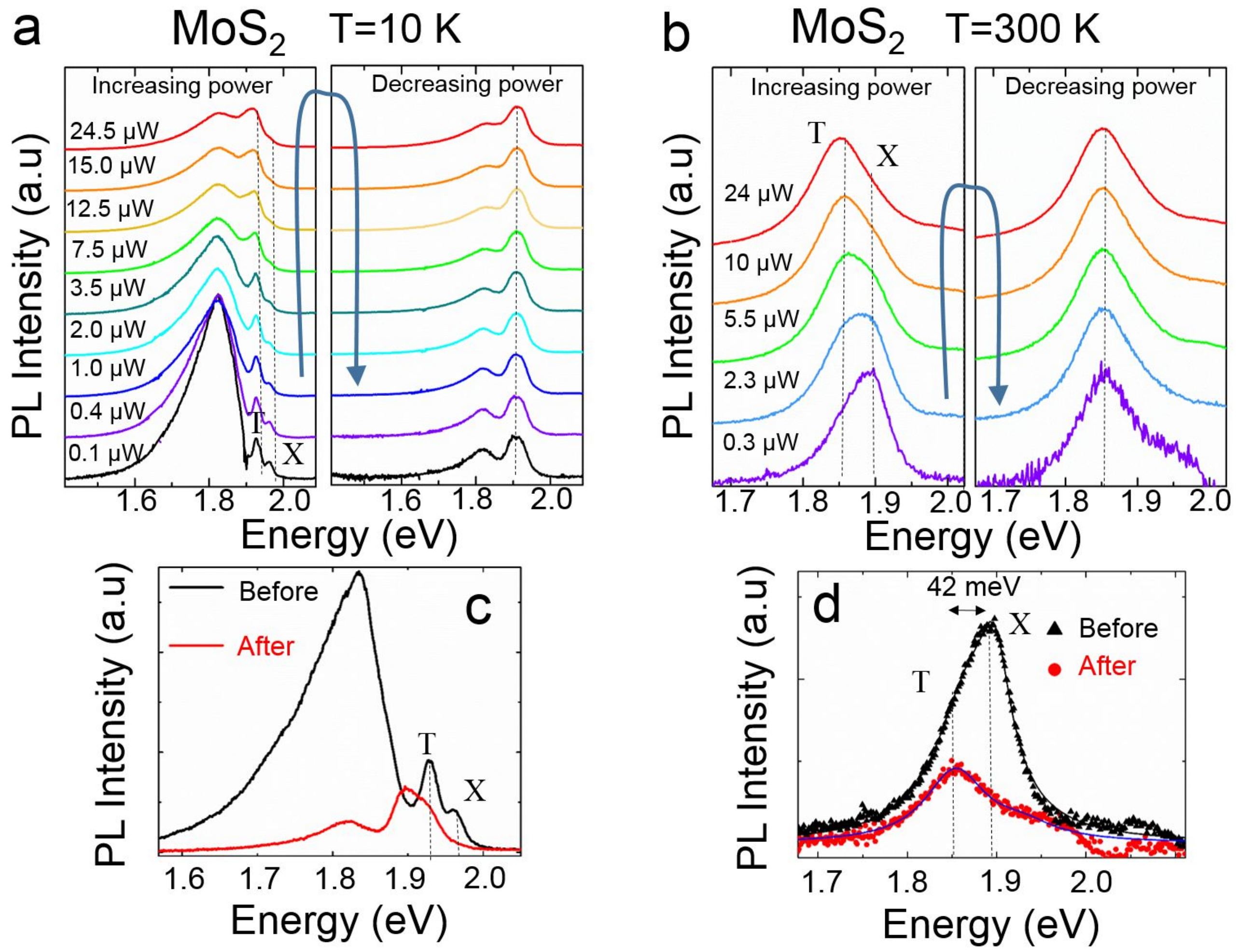}
\caption{\label{fig:fig4} \textbf{\underline{pulsed} Laser treatment of ML MoS$_2$} (a) PL spectrum of a MoS$_2$ ML at $T=10$ K as a function of \textit{average} power for a pulsed ($1.5$ ps) excitation at $400$ nm, revealing a hysteresis of the PL emission at low power created by the laser-induced doping of the ML. The spectra are labelled by \textit{average} power/$\mu$m$^2$.(b) Same as (a) but at T$=300$~K in vacuum. (c) PL spectrum at $T=10$ K and $0.1\;\mu$W excitation before (black curve) and after (red curve) exposure at $24.5$ $\mu$W for several minutes. (d) same as (c) but at $T=300$~K in vacuum. The hysteresis after laser exposure is also visible, with a redshift of the spectrum by $\sim 40$ meV when the PL is presumably dominated by the trion.  
}
\end{figure*}
\indent By exposing different regions of the MoS$_2$ ML to different laser powers, it is possible to control the $T/X$ ratio, and therefore, the doping of the ML. This is demonstrated in panel (c) of Fig.\ref{fig:fig3}, in which increasing the laser power between $10\;\mu$W and $200\;\mu$W allows to tune the $T/X$ ratio by almost one order of magnitude. In these experiments we have used different laser power exposure on different sample spots to control the T/X ratio locally. As shown for MoSe$_2$ MLs in Fig.~\ref{fig:fig1}a, an increase of the trion's spectral weight is accompanied by an increase of the trion's dissociation energy. In panel (d) of Fig.\ref{fig:fig3}, the PL of a MoS$_2$ ML after laser treatment probed subsequently with $0.4\;\mu$W excitation reveals high doping as inferred from the strong T emission (black curve). By illuminating the same region of the ML with an halogen lamp focused into a spot of $\sim 1\;\mu$m, the differential reflectivity of this doped region is obtained. The spectrum plotted in Fig.~\ref{fig:fig3}d (red curve)  indicates that the X transition is still the dominant absorption mechanism and that no significant shift of the X transition is observed with respect to undoped regions. This suggests that no significant band-gap renormalization is induced by the laser treatment. The same conclusion can be drawn for the MoSe$_2$ MLs studied, as the X peak PL energy (the optical band gap) in Fig.~\ref{fig:fig1}a,b and c does not change after the laser treatment. \\
\indent \textit{\textbf{Treatment of ML MoS$_2$ with \underline{pulsed} Lasers}.---} Finally, we demonstrate that a further increase of the laser excitation power completely quenches the X emission in MoS$_2$ MLs, as is shown in panel (a) of Fig.~\ref{fig:fig4}. For these more extreme experiments, a \textit{pulsed} excitation with 80~MHz repetition frequency at $400$ nm was used, in contrast to all experiments shown so far carried out with weak cw sources that can be found for example in commercial PL/Raman systems. At the lowest \underline{average} power used of 0.1~$\mu$W/$\mu$m$^2$ (\underline{peak} power of 600$\mu$W/$\mu$m$^2$), the PL spectrum presents X,T, and localized exciton emission. Increasing the average excitation power up to $\sim$25~$\mu$W/$\mu$m$^2$ results in a complete disappearance of the X line and a dramatic reduction of the localized exciton emission. Only a broad peak about 20~meV below the T energy is observed, accompanied by a small shoulder at lower energies. The 2 PL spectra of ML MoS$_2$ before and after laser treatment are totally different. When decreasing the average excitation power back to 0.1~$\mu$W/$\mu$m$^2$, a dramatic hysteresis of the PL spectrum is observed, as illustrated by panel (c) of Fig.\ref{fig:fig4}, similar to experiments using elevated laser power  in ML WS$_2$ \cite{Shang:2015a,He:2016a}. Please note that in many studies, the broad peak at $\sim 1.9$ eV has been attributed to the neutral exciton emission of MoS$_2$, and that valley-polarization experiments have been often performed with HeNe laser excitation at 1.96~eV, which corresponds to a perfectly-resonant excitation of the neutral exciton transition. These findings demonstrate that for pulsed excitations, a time-averaged power of only a few $\mu$W is enough to change the MoS$_2$ MLs optical properties in a non-reversible way. This has to be taken into account when analysing the complex physics probed in time-resolved PL measurements \cite{Korn:2011a,Lagarde:2014a}, pump-probe \cite{Mai:2014a,Wang:2013d,Wang:2014c,Kumar:2014a} and Kerr rotations experiments \cite{Yang:2015a,Plechinger:2014a}, which are often carried out in this excitation power regime. \\
\indent Many experiments on MoS$_2$ with pulsed or cw excitation are carried out at room temperature \cite{Amani:2015a}. We demonstrate the effect of the excitation laser on the optical spectrum also at T=300~K in vacuum conditions, as shown in panel (b) of Fig.\ref{fig:fig4}. A clear redshift of the PL peak position is observed and also a hysteresis of the PL at the lowest power used is evident (panel (d) of Fig.\ref{fig:fig4}). \\
\indent \textit{\textbf{Discussion.}---} 
There are several physical processes that can contribute to the modification of the Trion-to-neutral exciton PL emission ratio due to laser treatment. Behind this lie the different physical origins of excess carriers coming from doping of the TMD material, charges trapped at the ML-substrate interface and molecules on the ML surface. \\
\indent One possibility is that the laser induced changes are purely electronic i.e. due to optical ionization of impurities. These effects can have lifetimes from fractions of second to days \cite{Francinelli:2004a}. These physical processes were initially uncovered by Staebler and Wronski in hydrogenated amorphous SiO$_2$ \cite{Staebler:1977a}. One possibility is that the additional charges are optically created from defects in SiO$_2$ and are subsequently transferred to the TMD ML. In addition, charges trapped at defects could be optically activated in the TMD ML itself, as suggested by photoconductivity measurements \cite{Furchi:2014a}. A charge transfer from SiO$_2$ to the TMD monolayer could be suppressed by insertion of an h-BN layer \cite{Ju:2014a}, which might explain the absence of optical doping for this particular sample in Fig.~\ref{fig:fig2}c. \\
\indent Another possibility is that local heating results in defect formation/modification, as suggested by our substrate dependent studies. In Fig~\ref{fig:fig3} we see that the laser treatment does not modify the emission for ML MoSe$_2$ exfoliated onto h-BN or gold. Their thermal conductivities are much larger than that of SiO$_2$, as $\kappa(\text{SiO2-amorphous})\approx1-2~$W/m.K, $\kappa(\text{h-BN})\approx300~$W/m.K \cite{Jo:2013a} and $\kappa(\text{Au})\approx300~$W/m.K \cite{Langer:1997a}. This may indicate that the laser-induced doping of TMD MLs is thermally driven. In this case, the contact between the ML and a good thermal conductor may facilitate heat dissipation and as a consequence prevent the laser-induced doping of the ML. As the neutral exciton transition energy did not shift measurably for ML MoSe$_2$ (Fig.~\ref{fig:fig2}a) and ML MoS$_2$ (Fig.~\ref{fig:fig3}a) during laser treatment at T=4~K, strong heating effects are not detected in our experiments. Thermal conductivity is not the only difference between the substrate materials. For example, the atomic flatness of h-BN can also influence charge trapping processes and defect creation/propagation in the ML \cite{Mak:2012a,Wang:2015f}.\\
\indent Although the exact microscopic mechanisms still need to be understood, there are several important practical implications coming from our experiments on the modification of the optical properties of 2D materials following exposure to laser radiation.
For thorough optical studies MoX$_2$ ML samples should be investigated at very low laser power i.e. nW/$\mu$m$^2$. When investigating power dependence, hysteresis effects are very likely to occur if the maximum laser power used is too high. Our laser treatment technique could be used to locally pattern doped regions in the 2D crystal, possibly by using patterned substrates (SiO$_2$/Si versus h-BN). Our work shows that the neutral exciton emission of ML MoS$_2$ exfoliated on SiO$_2$ is at 1.96~eV, i.e. excitation with a HeNe laser is resonant with the neutral exciton transition. This will results in sharp, intense Stokes-lines from resonant Raman scattering \cite{Carvalho:2015a, Zhang:2015a} superimposed on the PL signal. The interplay between laser treatment and super-acid treatment might help in the future to identify how these two techniques influence different type of defects \cite{Amani:2015a,Cadiz:2016a}.  Experiments probing \textit{pristine} samples that need high laser power to generate enough signal will be difficult to compare with low power measurements such as white light reflectivity, as the optical properties will be modified.\\
\indent \textit{\textbf{Acknowledgements.}---} 
We thank ANR MoS2ValleyControl, ERC Grant No. 306719 and ITN SpinNANO for financial support. X.M. also acknowledges the Institut Universitaire de France. F.C. and P.R. thank the grant NEXT 
No ANR-10-LABX-0037 in the framework of the « Programme des Investissements d'Avenir". K.W. and T.T. acknowledge support from the Elemental Strategy Initiative
conducted by the MEXT, Japan and a Grant-in-Aid for Scientific Research on
Innovative Areas "Science of Atomic Layers" from JSPS. 
S.T. acknowledges support from National Science Foundation (DMR-1552220) and thanks INSA Toulouse for a visiting Professorship grant. 

\end{document}